\documentclass[prb,twocolumn,showpacs,floatfix,amsmath,amssymb,superscriptaddress]{revtex4}
\usepackage{latexsym}
\usepackage{graphicx}
\usepackage{times}
\usepackage{amsmath}
\usepackage{dcolumn}
\usepackage{latexsym,amsmath,amssymb,bm,euscript}
\newcommand{\bk}{{\bf k}}
\newcommand{\bd}{{\bm \delta}}
\newcommand{\cH}{{\cal H}}

\begin{document}

\title{Engineering a robust quantum spin Hall state in graphene via adatom deposition}

\author{Conan Weeks}
\affiliation{Department of Physics and Astronomy, University of British Columbia, Vancouver, BC, Canada V6T 1Z1}

\author{Jun Hu}
\affiliation{Department of Physics and Astronomy, University of California, Irvine, California 92697}

\author{Jason Alicea}
\affiliation{Department of Physics and Astronomy, University of California, Irvine, California 92697}

\author{Marcel Franz}
\affiliation{Department of Physics and Astronomy, University of British Columbia, Vancouver, BC, Canada V6T 1Z1}

\author{Ruqian Wu}
\affiliation{Department of Physics and Astronomy, University of California, Irvine, California 92697}

\date{\today}

\begin{abstract}
{The 2007 discovery of quantized conductance in HgTe quantum wells delivered the field of topological insulators (TIs) its first experimental confirmation. While many three-dimensional TIs have since been identified, HgTe remains the only known two-dimensional system in this class.  Difficulty fabricating HgTe quantum wells has, moreover, hampered their widespread use.  With the goal of breaking this logjam we provide a blueprint for stabilizing a robust TI state in a more readily available two-dimensional material---graphene.  Using symmetry arguments, density functional theory, and tight-binding simulations, we predict that graphene endowed with certain heavy adatoms realizes a TI with substantial band gap. For indium and thallium, our most promising adatom candidates, a modest $6\%$ coverage produces an estimated gap near 80K and 240K, respectively, which should be detectable in transport or spectroscopic measurements. Engineering such a robust topological phase in graphene could pave the way for a new generation of devices for spintronics, ultra-low-dissipation electronics and quantum information processing. 
}
\end{abstract}

\maketitle


\section{Introduction}

When certain physical properties of a system depend on global topology---and not on local details, such as disorder---then the system is said to realize a {\em topological phase}. By virtue of their universality, rooted in this independence on microscopic details, topological phases are both paradigmatically interesting and thought to possess seeds of important future technologies.  Until rather recently, experimental studies of topological phases were confined to the quantum Hall effect, seen in high-quality two-dimensional electron systems subjected to strong magnetic fields.  The advent of topological insulators (TIs), a class of two- and three-dimensional non-magnetic crystalline solids enjoying strong spin-orbit coupling\cite{MooreReview,KaneReview,QiReview}, opened a fascinating new chapter in the field.

Two-dimensional TIs are commonly referred to as quantum spin Hall (QSH) states\cite{MooreReview,KaneReview,QiReview} due to their similarity with integer quantum Hall liquids.  A defining signature of a QSH state is the existence of gapless edge states which are protected from elastic backscattering and localization by time-reversal symmetry $(\mathcal{T})$.  This feature, combined with the spin-filtered nature of the edge modes, renders such systems technologically very promising.  Furthermore, when in contact with a superconductor the edges are predicted to host Majorana fermions\cite{MajoranaQSH}, which play an important role in topological quantum computation schemes\cite{TQCreview}.  The QSH state was first predicted to arise in graphene\cite{KaneMele}, though later it was realized that, due to carbon's small atomic number, spin-orbit coupling is too weak to produce an observable effect under realistic conditions.  (First-principles calculations\cite{GrapheneSO1,GrapheneSO2,GrapheneSO3,GrapheneSO4,GrapheneSO5} on pristine graphene all predict sub-Kelvin gaps for this phase).  The QSH state was subsequently predicted in several other materials, including HgTe and InAs/GaSb quantum wells\cite{HgTeQSHtheory,QuantumWellQSH} and bilayer bismuth\cite{BismuthQSH}.  While landmark experiments indeed observed this phase in HgTe\cite{HgTeQSHexpt}, experimental activity on the QSH effect has remained limited by notorious practical difficulties associated with this material.  

Given the comparative ease with which graphene can be fabricated, it would be highly desirable to artificially enhance graphene's spin-orbit coupling strength to elevate the gap protecting the QSH state to observable levels. A practical means of doing so would pave the way to a new generation of QSH experiments accessible to a broad spectrum of researchers worldwide.  Previous work has explored the possibility that strong interactions\cite{Raghu,Weeks,Tami} or radiation\cite{Radiation1} might drive graphene into a robust topological phase.  Our goal in this manuscript is to explore a practical new route to this end.  Specifically, we will ask whether graphene can `inherit' strong spin-orbit coupling from a dilute concentration of heavy adatoms (whose innate spin-orbit coupling strength can be on the \emph{electron-Volt} scale) deposited randomly into the honeycomb lattice.

\begin{figure}
\includegraphics[width = 7.5cm]{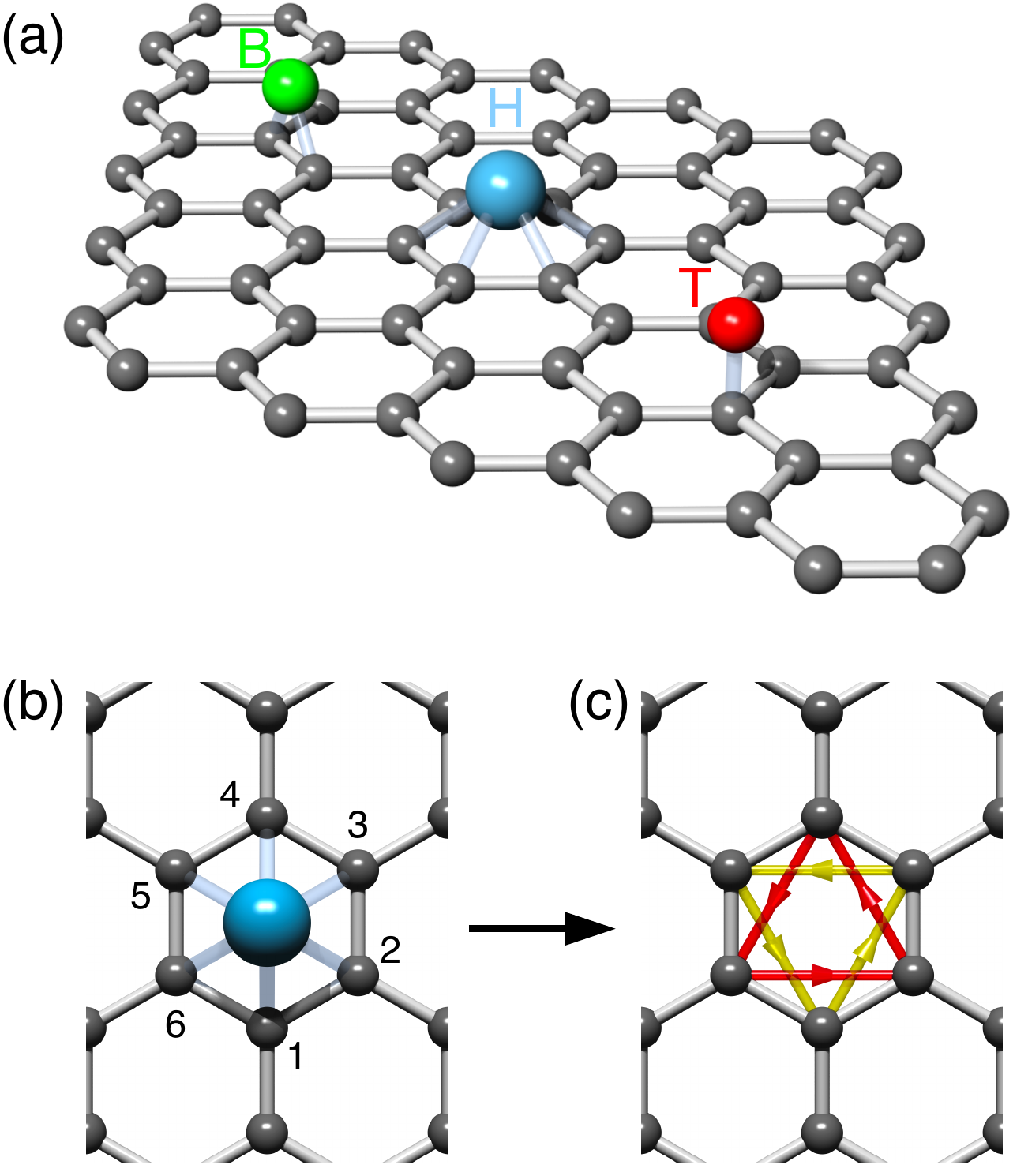}
\caption{{\bf Adatoms in graphene.} (a) Depending on the element, adatoms favor either the high-symmetry `bridge' (B), `hollow' (H), or `top' (T) position in the graphene sheet. (b) Detailed view of an H position adatom, which is best suited for inducing the intrinsic spin-orbit coupling necessary for stabilizing the topological phase.  The desired spin-orbit terms mediated by the adatom are illustrated in (c). Red and yellow bonds represent the induced second-neighbor imaginary hopping, whose sign is indicated by the arrows for spin up electrons. For spin down electrons the arrows are reversed.
}
\label{fig1}
\end{figure}
The basic principle underlying our proposal can be understood by considering processes in which an electron from graphene tunnels onto an adatom---whereupon it `feels' enormous spin-orbit coupling---and then returns to the graphene sheet.  Such processes in effect \emph{locally} enhance graphene's spin-orbit strength manyfold.  A viable proposal must, however, address numerous competing effects.    
For instance, adatoms often form local magnetic moments\cite{CohenAdatoms} which potentially spoil the time-reversal symmetry protecting the QSH effect.  Moreover, adatoms generically mediate both intrinsic \emph{and} Rashba spin-orbit coupling.  The latter is believed to be detrimental to the QSH phase\cite{KaneMele}, and previous work has indeed established that certain kinds of adatoms do generate substantial Rashba coupling in graphene that typically overwhelms the intrinsic contribution  \cite{ImpurityInducedSOC,AdatomSOC,QAHadatoms}.  (As Ref.\ \onlinecite{QAHadatoms} showed, however, magnetic adatoms inducing strong Rashba coupling can induce an interesting `quantum anomalous Hall' state in graphene.)  The adatoms may also favor competing, ordinary insulating states depending on their precise locations in the lattice.  And finally, since spin-orbit coupling is generated non-uniformly in graphene, the stabilization of a QSH phase even in an otherwise ideal situation is unclear \emph{a priori}.  

After an extensive search using tight-binding and first-principles analyses, we have found that two elements, indium and thallium, are capable of stabilizing a robust QSH state in graphene.  Neither element forms a magnetic moment, and although they do generate significant Rashba coupling, for symmetry reasons this remarkably does not suppress the QSH state.  We find that gaps many orders of magnitude larger than that predicted in pure graphene can form even with coverages of only a few percent; for example, at 6\% coverage indium yields a gap on the order of 100K, while for thallium the gap approaches room temperature.  These predictions revive graphene as a viable QSH candidate, and can be verified by probing the gap and associated edge states using spectroscopic and conductance measurements.

\section{Physics of a single heavy adatom}

To set the stage, let us first briefly review the Kane-Mele model\cite{KaneMele} describing pure, undoped graphene with spin-orbit coupling.  The Hamiltonian can be expressed as $H_{\rm KM} = H_t + H_{\rm so}$, where $H_t$ describes the usual nearest-neighbor hopping and $H_{\rm so}$ encodes intrinsic spin-orbit coupling.  In terms of operators $c^\dagger_{{\bf r}\alpha}$ that add electrons with spin $\alpha$ to site ${\bf r}$ of the honeycomb lattice and Pauli matrices $s^{x,y,z}$ that act on the spin indices, $H_t$ and $H_{\rm so}$ explicitly read
\begin{eqnarray}
  H_t &=& -t \sum_{\langle {\bf r r'}\rangle}(c^\dagger_{{\bf r}}c_{{\bf r'}} + h.c.)
  \label{Ht}\\
  H_{\rm so} &=& \lambda_{\rm so} \sum_{\langle\langle {\bf r r'}\rangle\rangle}(i \nu_{\bf r r'}c^\dagger_{{\bf r}}s^z c_{{\bf r'}} + h.c.).
  \label{Hso}
\end{eqnarray} 
Here and below, spin indices are implicitly summed whenever suppressed.
In Eq.\ (\ref{Hso}) the sum runs over second-nearest-neighbor lattice sites, and $\nu_{\bf r r'}$ are signs that equal $+1$ if an electron hops in the direction of the arrows in Fig.\ \ref{fig1}(c) and $-1$ otherwise.  Thus $H_{\rm so}$ describes `chiral' spin-dependent second-neighbor electron hopping.  When $\lambda_{\rm so} = 0$, the band structure exhibits the familiar gapless Dirac cones centered on momenta $\pm {\bf Q}$, resulting in semimetallic behavior.  Turning on $\lambda_{\rm so} \neq 0$ generates an energy gap\cite{KaneMele} $\Delta = 6 \sqrt{3}|\lambda_{\rm so}|$ at the Dirac points, transforming the system into a (very fragile\cite{GrapheneSO1,GrapheneSO2,GrapheneSO3,GrapheneSO4,GrapheneSO5}) QSH insulator.  Importantly, if mirror symmetry with respect to the graphene plane is broken, then Rashba coupling---which involves spin flips and thus breaks the U(1) spin symmetry enjoyed by $H$---will also be present\cite{KaneMele}.  Rashba coupling competes with the intrinsic spin-orbit term in pure graphene, and beyond a critical value closes the gap and destroys the QSH state.  

If heavy adatoms are to stabilize a more robust QSH phase in graphene, then at a minimum they should be non-magnetic (to preserve ${\mathcal{T}}$) and modify the physics near the Dirac points primarily by inducing large \emph{intrinsic} spin-orbit coupling.  The latter criterion leads us to focus on elements favoring the `hollow' (H) position in the graphene sheet indicated in Fig.\ \ref{fig1}(a).  Compared to the `bridge' (B) and `top' (T) positions, adatoms in the H position can most effectively mediate the spin-dependent second-neighbor hoppings present in the Kane-Mele model, while simultaneously avoiding larger competing effects such as local sublattice symmetry breaking generated in the T case.    

Since H-position adatoms generically reside on one side of the graphene sheet, they will clearly mediate Rashba spin-orbit coupling as well, leading to a potentially delicate competition.  If the adatoms' outer-shell electrons derive from $p$-orbitals, however, the induced intrinsic spin-orbit terms \emph{always} dominate over the induced Rashba interactions.  One can establish this key result by studying graphene with a single adatom of this type.  To model this setup, we employ operators $d_{m\alpha}$ for the adatom states, with $m = 0,\pm 1$ and $\alpha = \uparrow,\downarrow$ labeling the orbital and spin angular momentum quantum numbers.  The coupling of these orbitals to graphene is conveniently expressed in terms of the following operators:
\begin{equation}
  C_{m\alpha} = \frac{1}{\sqrt{6}}\sum_{j = 1}^6 e^{-i\frac{\pi}{3}m(j-1)} c_{{\bf r}_j\alpha},
  \label{Cm}
\end{equation}
where the sum runs over the six sites surrounding the adatom shown in Fig.\ \ref{fig1}(b).  Guided by symmetry (see Appendix A), we consider the following minimal Hamiltonian for this single-adatom problem:
\begin{eqnarray}
  H &=& H_{g} + H_{a} + H_{c}
  \label{Hadatom}
  \\
  H_g &=& H_t -\delta \mu\sum_{j = 1}^6 c^\dagger_{{\bf r}_j}c_{{\bf r}_j}
  \label{Hg}
  \\
  H_a &=& \sum_{m = 0,\pm 1} \epsilon_{|m|} d_{m}^\dagger d_{m} 
  + \Lambda_{\rm so}(d_{1}^\dagger s^z d_{1} - d_{-1}^\dagger s^z d_{-1})
  \nonumber \\
  &+& \sqrt{2}\Lambda'_{\rm so}(d_0^\dagger s^- d_{-1} + d_0^\dagger s^+ d_1 + h.c.)
  \label{Ha} \\
  H_c &=& -\sum_{m = 0,\pm 1} (t_{|m|} C_{m}^\dagger d_{m} + h.c.),
  \label{Hc}
\end{eqnarray}
with $s^\pm = (s^x \pm i s^y)/2$.
Here $H_g$ represents the nearest-neighbor hopping model of graphene supplemented by a chemical potential $\delta \mu$ for the six sites surrounding the adatom.  Physically, a non-zero $\delta \mu$ leads to an excess electron density at those sites, screening any net charge from the adatom.  Crystal field effects and spin-orbit coupling split the adatom $p$-orbitals through the $\epsilon_{|m|}$ and $\Lambda_{\rm so},\Lambda'_{\rm so}$ terms in $H_a$.  Finally, $H_c$ allows electrons to tunnel between the adatom and its neighboring carbon sites.  All couplings in $H$ are real except $t_{1}$, which is pure imaginary.  Note also that we have ignored the exceedingly weak spin-orbit terms that couple the electrons in graphene directly, as well as Coulomb interactions for the adatom.  Exclusion of the latter terms is justified for the non-magnetic adatoms we will consider below.  

Recalling that intrinsic spin-orbit coupling in graphene conserves the $S^z$ spin component while Rashba coupling does not, one can readily understand how the adatom mediates both kinds of interactions by viewing the tunnelings in $H_c$ perturbatively.  For example, when an electron hops onto the $m = 0$ orbital via $t_0$, flips its spin through the $\Lambda_{\rm so}'$ coupling, and then hops back to the honeycomb lattice via $t_1$, spin-orbit coupling of Rashba type is locally generated in graphene.  In sharp contrast to the situation for the Kane-Mele model, however, Rashba terms mediated in this fashion are \emph{irrelevant} for the low-energy physics.  This crucial feature can be understood by Fourier transforming $C_{0\alpha}$; remarkably, the components with Dirac-point momenta $\pm {\bf Q}$ vanish identically.  More physically, electrons near the Dirac points interfere destructively when hopping onto the $m = 0$ adatom orbital.  (See Appendix A and D for a complementary perspectives.)  We stress that this argument holds only for $p$-orbitals.  If the relevant adatom orbitals carry higher angular momentum, spin-flip processes which \emph{do} affect the Dirac points will generically appear.  The induced Rashba terms may still be subdominant, though whether this is the case depends on details of the Hamiltonian unlike the situation for $p$-orbitals.  

In contrast, $S^z$-conserving events whereby an electron hops both onto and then off of an $m = \pm 1$ orbital via $t_1$ locally mediate the desired \emph{intrinsic} spin-orbit interactions.   
No obstruction exists for tunneling onto the $m = \pm 1$ orbitals via $t_1$, so these processes \emph{can} effectively modify the physics near the Dirac points.  It is important to note, however, that $t_1$ hopping mediates additional couplings as well, so the dominance of these spin-orbit terms remains unclear.  Nevertheless, given the high symmetry preserved by the H-site adatom, none of the conventional broken-symmetry gapped phases of graphene---such as charge-density wave or `Kekule' orders---are obviously favored here, so it is reasonable to expect these additional terms to play a relatively minor role.  We explicitly confirm this intuition in the multi-adatom situation, to which we now turn.

\section{Periodic adatom configurations}

As a first step in understanding the multi-adatom case, we examine a periodic system with one adatom residing in a large $N\times N$
supercell. This situation allows us to utilize density functional theory (DFT) to ascertain suitable heavy elements and obtain a
quantitative understanding of their effects on graphene.  (See Appendix B for computational details.) To ensure large spin-orbit
coupling, we focused on elements in rows five and six of the periodic table, including In, Sn, Sb, Te, I, La, Hf, Pt, Au, Hg, Tl, Pb and
Bi. For each element, we calculated the total energy in the three adsorption geometries shown in Fig.\ 1(a) along with the adatom's spin
moment. Our calculations reveal that two elements---indium (atomic number $Z = 49$) and thallium ($Z = 81$)---satisfy our criteria of both
favoring the high-symmetry H position \emph{and} being non-magnetic.  Furthermore, both elements exhibit partially filled $p$-shells,
ensuring that the Rashba coupling they mediate in graphene is benign at the Dirac points.  

\begin{figure*}
\includegraphics[width = 15 cm]{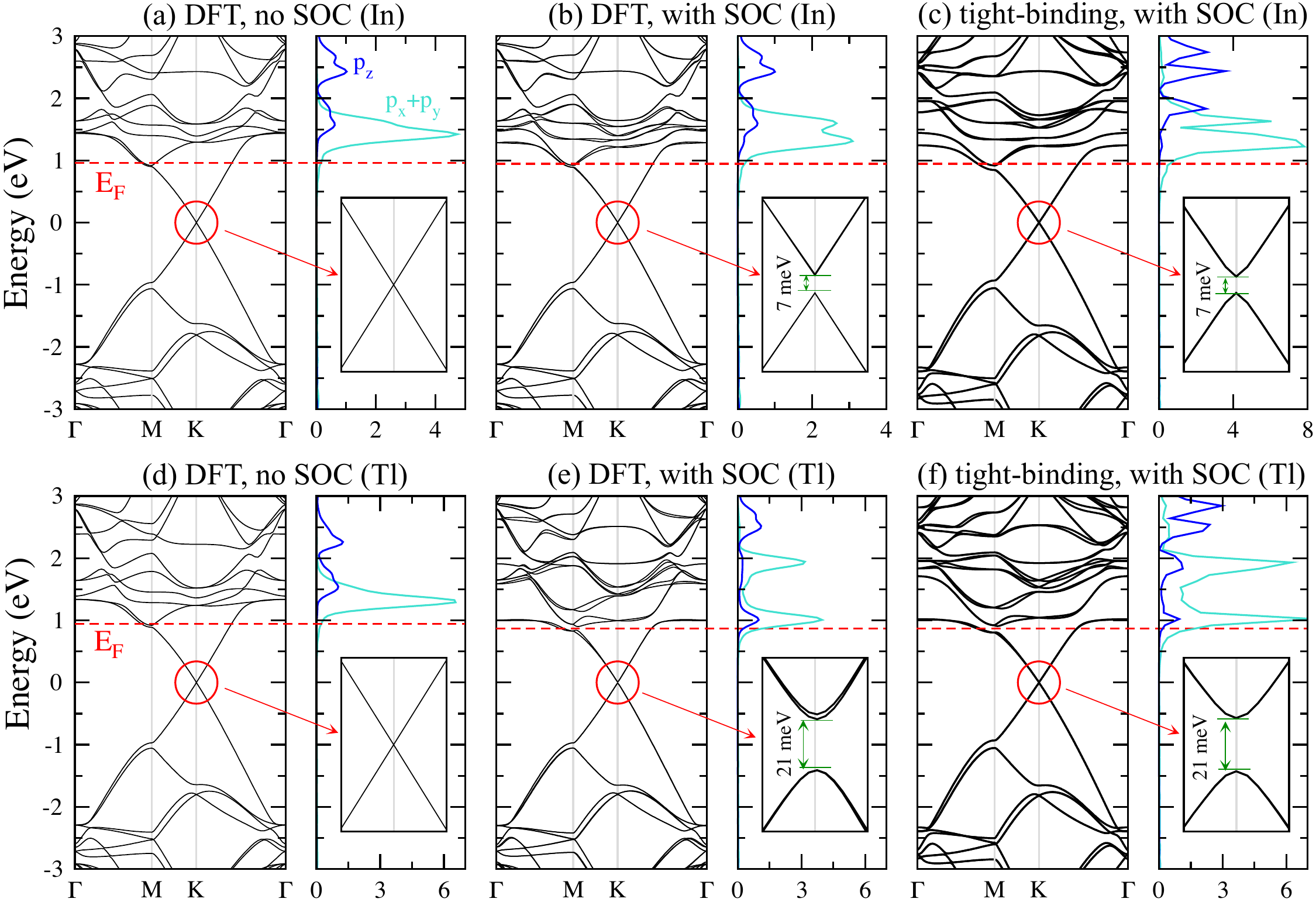}
\caption{{\bf Band structure and the adatom local density of states (LDOS).}  All data correspond to one adatom in a $4\times4$ supercell,
with the upper row corresponding to indium and the lower row corresponding to thallium. The left panels in (a) and (d)
correspond to the band structure and LDOS computed using DFT \emph{without} spin-orbit coupling.  The horizontal dashed lines indicate the
Fermi level ($E_F$), which shifts due to electron-doping from the adatoms. Insets zoom in on the band structure near the K point within
an energy range $-35$ to 35meV, showing that without spin-orbit interactions neither indium nor thallium open a gap at the Dirac points. The
central panels in (b) and (e) are the corresponding DFT results including spin-orbit coupling.  Remarkably, in the indium case a
gap of 7meV opens at the Dirac point, while with thallium the gap is larger still at 21meV.  Finally, the right panels in (c) and
(f) were obtained using the tight-binding model described in the text. }
\label{fig2}
\end{figure*}

It is instructive to first examine the electronic properties of indium on graphene in a $4\times 4$ supercell, \emph{without} spin-orbit
coupling. As Fig.\ \ref{fig2}(a) illustrates, the Dirac cones characteristic of pure graphene indeed remain massless---despite the reduced
translation symmetry, conventional gapped phases are not stabilized here, consistent with the intuition developed in the single-adatom case
above.  Indium does, however, electron-dope graphene and shifts the Fermi level $E_F$ to 0.95 eV above the Dirac points. From the adatom's
local density of states (LDOS) displayed in Fig.\ \ref{fig2}(a), one can see that indium's $5p$ orbitals lie almost entirely above
$E_F$, implying that the $5p$ electron in neutral indium nearly completely transfers to graphene.  (The charge of an indium adatom is $+0.8
e$ from the Bader charge division scheme.) Note that the relatively diffuse $p_z$ LDOS indicates that this orbital hybridizes more strongly
with graphene compared to the $p_{x,y}$ orbitals.  Replacing indium with thallium, again without spin-orbit coupling, leads to the band
structure and LDOS shown in Fig.\ \ref{fig2}(d).  Clearly the electronic structure is modified very little by this substitution;
importantly, the Dirac cones remain massless with thallium as well.  

\begin{figure}[t]
\includegraphics[width = 8.5cm]{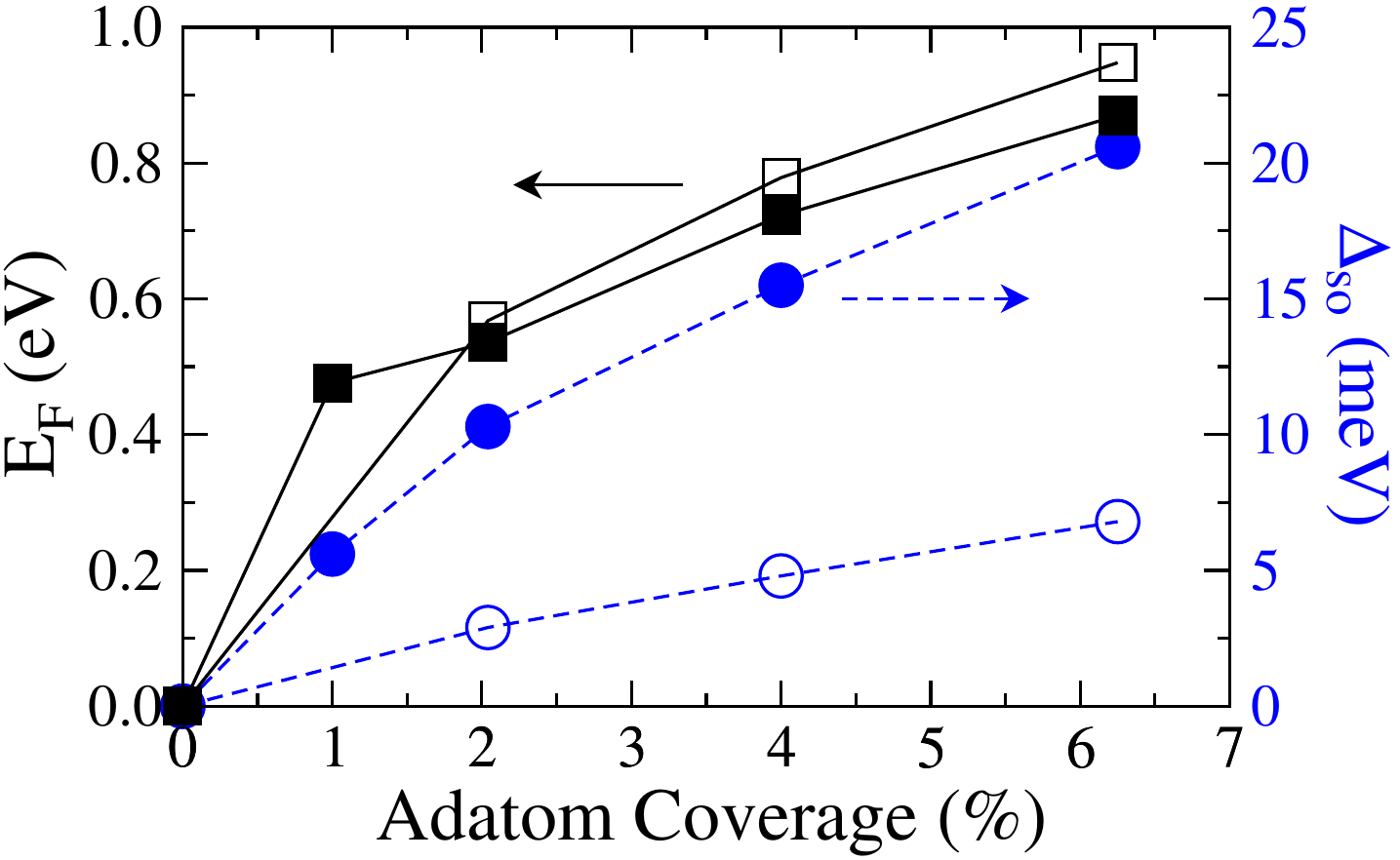}
\caption{{\bf Coverage Effects.}  Spin-orbit-induced gap $\Delta_{\rm so}$ opened at the Dirac points and Fermi level $E_F$ (measured relative to the center of the gap) for
different indium and thallium adatom coverages. The open and filled symbols represent data for indium and thallium, respectively.}
 \label{fig3}
\end{figure}

Thus any gap opening at the Dirac points must originate from spin-orbit coupling.  Fig.\ \ref{fig2}(b) displays the band structure
and LDOS for spin-orbit-coupled indium on graphene.  Note the sizable spin-orbit splitting in the LDOS for the $p_{x,y}$ orbitals.  More remarkably, a gap $\Delta_{\rm so}\approx 7$meV now appears at the Dirac points, which already exceeds the
spin-orbit-induced gap in pure graphene\cite{GrapheneSO1,GrapheneSO2,GrapheneSO3,GrapheneSO4,GrapheneSO5} by several orders of magnitude. 
The analogous results for thallium---whose atomic mass is nearly twice that of indium---are still more striking.  As Fig.\ \ref{fig2}(e)
illustrates, $p$-orbital splittings of order 1eV are now evident in the LDOS, and a gap $\Delta_{\rm so}\approx 21$meV opens at the
Dirac points.  We emphasize that these results apply for adatom coverages of only 6.25\%. To explore the dependence of $\Delta_{\rm so}$ on the
adatom coverage, we additionally investigated systems with one adatom in $5\times5$, $7\times7$ and $10\times10$ supercells. (For $N\times
N$ supercells with $N$ a multiple of 3, the Dirac points reside at zero momentum and can thus hybridize and gap out even without spin-orbit
coupling.  We thus ignore such geometries.)  The values of $\Delta_{\rm so}$ along with the Fermi level $E_F$ computed for the coverages we
studied appear in Fig.\ \ref{fig3}.  Quite naturally, the gap decreases as one reduces the coverage, as does the Fermi level since fewer electrons are donated to graphene at lower adatom concentrations.  It is worth highlighting that the gap decreases sublinearly and that a sizable $\Delta_{\rm so} \approx 10$meV remains even with a mere 2.04\% thallium concentration.

Since spin-orbit coupling clearly underlies the gap onset, it is tempting to associate the insulating regime opened by the adatoms with a QSH state.  To verify this expectation, we appeal to tight-binding simulations of the $4\times4$ supercell geometry.  We model the system by a Hamiltonian $H_{4\times4}$ consisting of uniform hopping for graphene (including weak second- and third-neighbor tunneling) supplemented by the additional local terms in Eq.\ (\ref{Hadatom}) at each adatom.  Figure \ref{fig2}(c) displays the tight-binding band structure and adatom LDOS for indium with the following parameters: first, second, and third-neighbor graphene hoppings $t = 2.82$eV, $t' = 0.22$eV, and $t'' = 0.2$eV; $\delta\mu = 0.5$eV; $\epsilon_0 = 2.5$eV, $\epsilon_1 = 1.8$eV; $t_0 = 2$eV, $it_1 = 0.95$eV; and $\Lambda_{\rm so} = \Lambda_{\rm so}' = 0.1$eV.  The corresponding data for thallium appear in Fig.\ \ref{fig2}(f); parameters are the same as for indium, except for $\epsilon_1 = 1.9$eV and $\Lambda_{\rm so} = \Lambda_{\rm so}' = 0.31$eV.  The good agreement with DFT provides strong evidence that the underlying physics is well-captured by our tight-binding model.  With this in hand, one can then demonstrate the topological nature of the insulating regime by showing that the Dirac point gap remains non-zero upon adiabatically deforming the Hamiltonian to the Kane-Mele model.  This can be achieved by defining a new Hamiltonian $H(\lambda) = (1-\lambda)H_{4\times4} + \lambda H_{\rm KM}$ that interpolates between our adatom model at $\lambda = 0$ and the Kane-Mele model at $\lambda = 1$.  For either indium or thallium parameters, the Dirac point gap never closes as $\lambda$ varies from 0 to 1 (see Appendix C for details).  Thus either element can indeed stabilize a robust QSH state in graphene, at least when periodically arranged.
%

\section{Randomly distributed adatoms}

\begin{figure*}[t]
\includegraphics[width = 14.0cm]{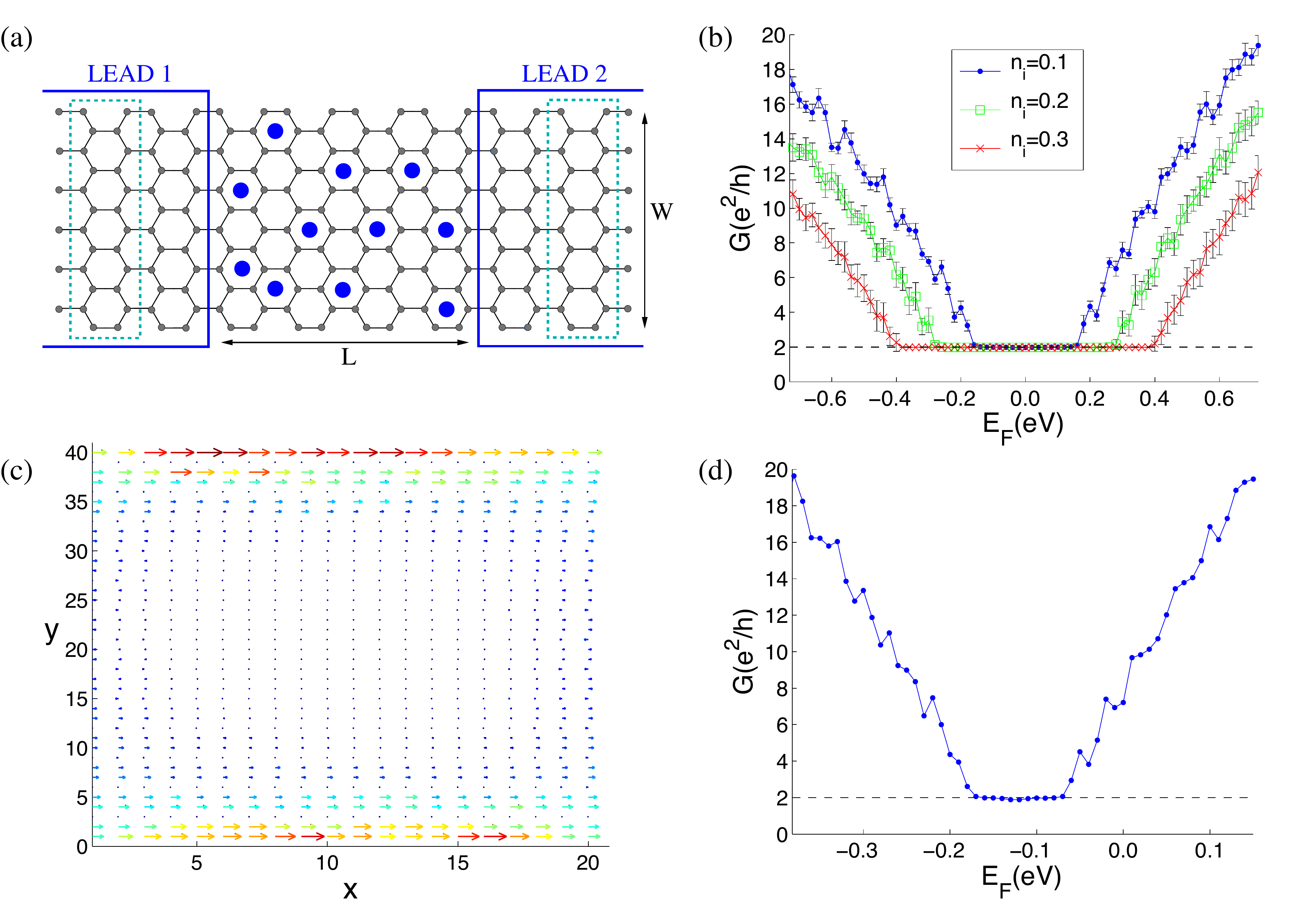}
\caption{{\bf Transport.} (a) Configuration used to
determine the two-terminal conductance. Semi-infinite clean
graphene leads are connected to the sample, which has adatoms
distributed randomly across it. The length $L$ of the system is set by
the number of columns (indicated by dotted lines). The width $W$
refers to the number of sites along either edge of the figure,
corresponding to the number of unit cells of an armchair nanoribbon
required to create an individual column.  (b) Conductance $G$ as a
function of the Fermi energy $E_F$, averaged over 40 independent random adatom
realizations at different concentrations $n_i=0.1,0.2,0.3$ for a
system of size $W=80$ and $L=40$ with $\lambda_{\rm so}=0.1t$. (c)
Current distribution across a sample of size $W=40$ and $L=20$ at
$n_i=0.2$, $\lambda_{\rm so}=0.1t$ and $E_{F}=0.15 eV$. The magnitude
of the current is represented by both the arrow size and color.  (d)
Conductance for the largest simulated system size using realistic
parameters for thallium adatoms ($\lambda_{\rm so}=0.02t$ and $\delta
\mu=0.1t$) estimated from DFT data. Here $W=200$, $L=100$, and the coverage is $n_i=0.15$. In all panels we take $t=2.7$eV.}
\label{fig4}
\end{figure*}

In an experiment adatoms will likely occupy random H-site locations in a graphene sheet that is of course not pristine. It is therefore important to understand the impact of such randomness and disorder on
the stability of the topological phase that was found above in pure graphene with periodic adatoms. 
To address general features of the problem in a way that reduces its computational complexity, we will describe the disordered, random-adatom system using the following graphene-only Hamiltonian,
\begin{equation} \label{hran1}
H=H_t+\sum_I\delta H_I-\sum_{\bf r}\delta\mu_{\bf r}c^\dagger_{{\bf r}}c_{{\bf r}}.
\end{equation}
Here $\delta\mu_{\bf r}$ represents a random on-site potential arising, \emph{e.g.}, from the substrate (and not the adatoms).  In the second term, $I$ labels the random plaquettes occupied by adatoms and 
\begin{eqnarray} \label{hran2}
\delta H_I&=&-\delta \mu\sum_{{\bf r}\in I} c^\dagger_{{\bf r}}c_{{\bf r}}
+\lambda_{\rm so} \sum_{\langle\langle {\bf r r'}\rangle\rangle\in I}(i \nu_{\bf r r'}c^\dagger_{{\bf r}}s^z c_{{\bf r'}} + h.c.) \nonumber\\
&&+i\lambda_R  \sum_{ {\bf r,r'}\in I}c^\dagger_{{\bf r}}({\bf s}\times{\bf d}_{\bf r r'})_z  c_{{\bf r'}}
\end{eqnarray}
with  ${\bf d}_{\bf r r'}={\bf r}- {\bf r'}$.  The first term in $\delta H_I$ describes the chemical potential that screens charge from the adatoms, while the last two terms capture the local intrinsic ($\lambda_{so}$) and Rashba ($\lambda_R$) spin-orbit couplings induced by electrons
hopping from graphene to an adatom and back, as discussed in Sec.\
II. Formally, these spin-orbit terms can be derived by
integrating out perturbatively the adatom degrees of freedom in Hamiltonian
(\ref{Hg}-\ref{Hc}); see Appendix D. It is important to notice that unlike the conventional nearest-neighbor Rashba term considered
in Ref.\ \onlinecite{KaneMele} for pristine graphene, the adatom-generated
Rashba term connects {\em all} sites in the hexagon. Such a `hexagon
Rashba term' has the property of vanishing at the Dirac points as
already argued in Sec.\ II.  The adatom also induces other symmetry-allowed terms,
such as further-neighbor spin-independent
hoppings, which we disregard because they are either weak or do
not lead to qualitative changes in the results reported
below.

To study the robustness of the QSH phase in the presence of random
adatoms and disorder we probe for the signature
gapless edge states by simulating the classic two-terminal conductance
measurement in the geometry of Fig.\ \ref{fig4}(a). Within the Landauer-B\"uttiker formalism \cite{landauer1,buttiker1} we employ the
recursive Green's function method \cite{bruno1} to evaluate the 
conductance $G$ of a length-$L$, width-$W$ graphene strip as we vary the Fermi energy.  Consider first the simplest random-adatom setup with $\delta\mu_{\bf r} = 0$ (corresponding to an otherwise clean graphene sheet), $\delta\mu = 0$, and $\lambda_{R} = 0$.  
Figure \ref{fig4}(b) illustrates the conductance for this case at several adatom concentrations $n_i$, and model
parameters indicated in the caption.  A $2e^{2}/h$ plateau
with width proportional to $n_i$ clearly emerges, strongly suggesting the onset of a bulk mobility gap and quantized ballistic conduction by edge states.  
This picture is corroborated by Fig.\ \ref{fig4}(c), which displays the current
distribution for $E_F=0.15eV$ across a smaller system size 
(chosen for clarity). 

These results establish that in principle adatoms need not be periodic to stabilize a QSH phase, though the parameters employed were unrealistic.  To make contact with our DFT results, we also considered a system with
$\lambda_{\rm so}=0.02t$ and $\delta \mu=0.1t$, which would yield a
similar bulk mobility gap to that seen in our thallium simulations if the adatom coverages were the same. 
While finite-size effects prevent us from studying the low-adatom coverages considered in Sec.\ III, Fig.\ 4(d) shows that for $n_i = 0.15$ a robust conductance plateau indeed persists for these more realistic parameter values.  (At much lower coverages the width of the edge states approaches the system sizes we are able to simulate.)  We have also studied the effects of the hexagon Rashba term and residual on-site potential disorder on the
stability of the QSH phase (see Appendix E for details).  Even a relatively large $\lambda_R\sim 2\lambda_{\rm so}$ has a weak effect on the width of the conductance plateau, as expected given our analysis in Sec.\ II.  With uncorrelated on-site disorder $\delta \mu_{\bf r}$, the topological phase is also remarkably robust---$\delta\mu_{\bf r}$ can fluctuate on the scale of $t$ while degrading the plateau only marginally.  The topological phase is more sensitive, however, to correlated disorder which is likely more relevant experimentally.  In this case the conductance plateau survives only when $\delta \mu_{\bf r}$ varies on a scale smaller roughly than the mobility gap for the clean case.  To observe the QSH phase experimentally, the chemical potential should therefore fluctuate on scales smaller than the $\sim10$meV gaps we predicted here. While this might be difficult to achieve with graphene on standard SiO$_2$ substrate, recent experiments using hexagonal boron nitride substrates\cite{hBN} show disorder energy scales as low as 15K, which should be sufficient to observe the QSH gap.

\section{Relation to previous work}

Since numerous recent studies have explored the possibility of employing adatoms to control graphene's electronic properties, here we briefly comment on the distinctions between our proposal and related earlier works.  Castro Neto and Guinea\cite{ImpurityInducedSOC} previously argued that graphene's spin-orbit strength can be greatly enhanced by adatom deposition, though the physical mechanism and nature of induced spin-orbit coupling is quite different from what we considered here.  The enhancement discussed in Ref.\ \onlinecite{ImpurityInducedSOC} stemmed from structural changes in the honeycomb lattice induced by a T-position adatom (such as hydrogen whose innate spin-orbit coupling is weak).  We explicitly neglected such contributions, as our DFT simulations demonstrated that the adatoms we considered only weakly affect the carbon-carbon bonds.  Perhaps even more importantly, the effect was discussed solely in the context of enhanced \emph{Rashba} spin-orbit interactions, which do not stabilize a QSH state in graphene.  

More recently, Abdelouahed \emph{et al}.\cite{AdatomSOC} examined the influence of heavy Au adatoms on graphene's electronic structure using first-principles calculations.  These authors considered dense coverages, with one T-position adatom residing above each carbon site on graphene's A sublattice, and found substantial enhancement of graphene's Rashba coupling to values of order 10meV.  Again, this type of induced spin-orbit coupling is not suitable for driving a QSH state, and in any case this geometry is very likely to produce only a topologically trivial band insulator due to explicit breaking of sublattice symmetry by the adatoms.  (Lighter Ni adatoms were also studied, but were found to modify graphene's spin-orbit coupling only very weakly.)  

Closest in spirit to our study is the very interesting work of Qiao \emph{et al}.\cite{QAHadatoms}, who considered the possibility that adatoms can stabilize another topological phase in graphene---a quantum anomalous Hall state.  Unlike the QSH phase, this state breaks time-reversal symmetry and exhibits chiral edge modes leading to a non-trivial Hall conductivity as in the integer quantum Hall effect.  Thus the requirements for stabilizing a quantum anomalous Hall phase in graphene are quite different.  Qiao \emph{et al}. showed that this can be accomplished by a combination of strong Rashba spin-orbit coupling and an exchange-induced Zeeman field.  They further demonstrated via first-principles calculations that 6.25\% coverage of periodically arranged iron adatoms (which form magnetic moments) provide both ingredients and generate a quantum anomalous Hall phase protected by a 5.5meV gap.  Note that while iron favors the H site like indium and thallium, its outer-shell electrons derive from $d$ rather than $p$ orbitals.  Thus our argument for the irrelevance of Rashba spin-orbit coupling at the Dirac points does not apply to iron adatoms, consistent with the numerical findings of Ref.\ \onlinecite{QAHadatoms}.  Although the gap produced by iron is smaller than that produced by indium and especially thallium at the same coverage, a promising feature of their proposal is that iron adatoms do not alter the Fermi level.

We also wish to point out a connection of our work to `topological Anderson insulators'---systems which are non-topological in the clean limit but acquire non-trivial topology when disordered.  Such states were first predicted in Ref.\ \onlinecite{TopologicalAndersonInsulator1} and explored further in several subsequent studies\cite{TopologicalAndersonInsulator2,TopologicalAndersonInsulator3,TopologicalAndersonInsulator4,TopologicalAndersonInsulator5}.  The robust QSH state that can emerge upon randomly depositing dilute adatoms onto pure graphene (which by itself is a non-topological semimetal) can be viewed as an example of a topological Anderson insulator.  Note, however, that this categorization is somewhat loose, as this phase in fact smoothly connects to the topological phase of the pure Kane-Mele model\cite{KaneMele}, which has no disorder (see also Ref.\ \onlinecite{TopologicalAndersonInsulator5}).  Nevertheless, this viewpoint is useful in that it highlights the remarkable fact that disorder, with suitable properties, can provide a practical avenue to generating topological phases.

\section{Discussion}

In this paper we have demonstrated that dilute heavy adatoms---indium and thallium in particular---are capable of stabilizing a robust QSH state in graphene, with a band gap exceeding that of pure graphene by many orders of magnitude.  Aside from fabrication ease, another virtue graphene possesses as a potential QSH system is the large number of experimental probes available in this material because the electrons are not buried in a heterostructure.  Angle resolved photoemission spectroscopy (ARPES), for instance, 
actually benefits from electron doping by the adatoms, and thus could be employed to detect the onset of the Dirac point gap at arbitrary coverages.  Scanning tunneling microscopy (STM) would be similarly well-suited for probing the bulk gap, the spatial structure of the LDOS around an adatom, and even the topologically protected gapless edge modes.  

Detecting the edge states via transport will be more challenging, since this would require repositioning the Fermi level inside of the bulk gap.  Nevertheless, the technology for doping graphene by the requisite amount does exist (see, \emph{e.g.}, Refs.\ \onlinecite{GrapheneGating1,GrapheneGating2,GrapheneGating3,FerroelectricSubstrate,SolidElectrolyte}). Although the most commonly used ion-liquid techniques \cite{GrapheneGating1,GrapheneGating2,GrapheneGating3} work best near room remperature, recently developed approaches employing ferroelectric substrates\cite{FerroelectricSubstrate} and solid polymer electrolytes\cite{SolidElectrolyte} show comparable results and remain applicable down to cryogenic temperatures. The requisite doping
may also be achievable by introducing additional high-electronegativity adatoms that absorb electrons. A quantum spin Hall phase in an appropriately doped graphene sample can be detected through the classic two-terminal conductance measurement akin to the historic observation performed in HgTe quantum wells\cite{HgTeQSHexpt}. A more complete characterization of the topological phase can be obtained through non-local edge state\cite{Roth1} and magnetotransport measurements which provide unambiguous evidence for topologically protected helical edge states \cite{Tkachov1,Maciejko1}.

Fulfilling Kane and Mele's original vision by artificially turning graphene into a strong spin-orbit system could have remarkable technological implications.  Among these, the possibility of employing the protected edge states for realizing topological superconductivity, Majorana fermions, and related phenomena such as non-Abelian statistics and `fractional Josephson effects' are particularly tantalizing\cite{MajoranaQSH}.  Interestingly, one of the primary requirements for observing these phenomena---generating a superconducting proximity effect in graphene---has already been achieved by numerous groups\cite{GrapheneProximityEffect1,GrapheneProximityEffect2,GrapheneProximityEffect3,GrapheneProximityEffect4}.  These exciting possibilities, along with potential spintronics and low-dissipation devices, provide strong motivation for pursuing the line of attack we introduced for stabilizing a robust QSH state in graphene.

\acknowledgments{It is a pleasure to acknowledge helpful discussions with Jim Eisenstein, Ilya Elfimov, Josh Folk, Erik Henriksen, Roland Kawakami, Shu-Ping Lee, Gil Refael, Shan-Wen Tsai, and Amir Yacoby.  J.A. gratefully acknowledges support from the National Science Foundation through grant DMR-1055522. Work at UBC was supported by NSERC and CIfAR.}

\appendix

\section{Discussion of the effective Hamiltonian for graphene with a single adatom}

In this section we will consider a sheet of graphene located in the $(x,y)$ plane, with a single adatom sitting at the H position (recall Fig.\ 1 of the main text).  An important feature of such an adatom is the high symmetry that it preserves compared to the cases where the adatom resides at the B or T positions.  In particular, the following symmetries remain here: $\pi/3$ rotation ($R_{\pi/3}$) and $x$-reflection ($R_x$) about the adatom site, as well as time-reversal ($\mathcal{T}$).  Note that since the adatom is displaced vertically out of the $(x,y)$ plane, mirror symmetry with respect to the graphene plane ($R_z$) is broken.  It will nevertheless be useful below to understand how operators transform under $R_z$, since this will give us insight as to which couplings mediate intrinsic versus Rashba spin-orbit coupling.  

We will continue to assume that $p$-orbitals form the active states for the adatom.  Also as in the main text, we employ operators $d^\dagger_{m\alpha}$ which add electrons with spin $\alpha$ and orbital angular momentum quantum number $m = 0,\pm 1$ to the adatom, and operators $c^\dagger_{{\bf r}\alpha}$ which add electrons with spin $\alpha$ to site ${\bf r}$ of graphene's honeycomb lattice.  These operators transform under $R_{\pi/3}$ according to
\begin{eqnarray}
  R_{\pi/3}: ~~~~~~~~~~ c_{\bf r} &\rightarrow& e^{i\frac{\pi}{6}s^z}c_{\bf r'}
  \\
  R_{\pi/3}: ~~~~~~~~~~   d_m & \rightarrow & e^{i\frac{\pi}{6}s^z}e^{i\frac{m\pi}{3}}d_m
\end{eqnarray}
where ${\bf r}'$ corresponds to ${\bf r}$ rotated by $\pi/3$.  Note that here and below, both $c_{\bf r}$ and $d_m$ are considered to be two-component objects, with the upper and lower components corresponding to spin up and down, respectively.  Similarly, under $R_x$ we have
\begin{eqnarray}
  R_{x}: ~~~~~~~~~~ c_{(x,y)} &\rightarrow& e^{i\frac{\pi}{2}s^x}c_{(-x,y)}
  \\
  R_{x}: ~~~~~~~~~~ d_{m} &\rightarrow& e^{i\frac{\pi}{2}s^x}d_{-m}.
\end{eqnarray} 
Time-reversal, which is antiunitary, sends 
\begin{eqnarray}
  \mathcal{T}: ~~~~~~~~~~ c_{\bf r} &\rightarrow& e^{i\frac{\pi}{2}s^y}c_{\bf r}
  \\
  \mathcal{T}: ~~~~~~~~~~ d_{m} &\rightarrow& (-1)^m e^{i\frac{\pi}{2}s^y}d_{-m}.
\end{eqnarray}
Note that the $(-1)^m$ factor above arises because one conventionally defines spherical harmonics such that $[Y_{\ell}^m]^* = (-1)^m Y_{\ell}^{-m}$.  
Finally, the operators transform under $R_z$ as
\begin{eqnarray}
  R_z: ~~~~~~~~~~ c_{\bf r} &\rightarrow& -e^{i\frac{\pi}{2}s^z}c_{{\bf r}}
  \\
   R_z: ~~~~~~~~~~ d_{m} &\rightarrow& (-1)^{1-m}e^{i\frac{\pi}{2}s^z}d_{m}
\end{eqnarray}
In the last equation, the factor of $(-1)^{1-m}$ reflects properties of spherical harmonics under $z\rightarrow -z$.

Our aim now is to construct a minimal tight-binding model for graphene with a single H-position adatom.  Following the main text, we write the Hamiltonian as $H = H_g + H_a + H_c$, where $H_g$ involves only the carbon degrees of freedom, $H_a$ involves only the adatom degrees of freedom, and $H_c$ couples the two.  Consider first $H_a$.  The most general quadratic form for the adatom Hamiltonian can be written
\begin{eqnarray}
  H_a = \sum_{\alpha = 0,x,y,z}\sum_{m,m' = -1}^1 u^\alpha_{m,m'}d_m^\dagger s^\alpha d_{m'},
\end{eqnarray}
where $s^0$ is the identity, $s^{x,y,z}$ are Pauli matrices, and $u^\alpha_{m,m'} = [u^\alpha_{m',m}]^*$ due to Hermicity.  
Using the transformation rules obtained above, one can show that symmetry strongly constrains the allowed $u^\alpha_{m,m'}$ terms.  For example, in the $\alpha = 0$ sector, $R_{\pi/3}$ implies that $u^0_{m,m'}$ must vanish except when $m = m'$, while $R_x$ further requires that $u^0_{m,m} = u^0_{-m,-m}$.  (Time-reversal constrains these couplings no further.)  The $\alpha = x,y,z$ terms can be similarly constrained, leading to the expression
\begin{eqnarray}
H_a &=& \sum_{m = 0,\pm 1} \epsilon_{|m|} d_{m}^\dagger d_{m} 
+ \Lambda_{\rm so}(d_{1}^\dagger s^z d_{1} - d_{-1}^\dagger s^z d_{-1})
  \nonumber \\
  &+& \sqrt{2}\Lambda'_{\rm so}(d_0^\dagger s^- d_{-1} + d_0^\dagger s^+ d_1 + h.c.)
\end{eqnarray}
provided in the main text.  

Let us elaborate on the physics here in greater detail.  If the adatom were in isolation, continuous rotation symmetry would further restrict $\epsilon_{|m|}$ to be independent of $m$ and require that $\Lambda_{\rm so}' = \Lambda_{\rm so}$.  In this case the $\Lambda_{\rm so}, \Lambda_{\rm so}'$ terms above simply encode an isotropic ${\bf L}\cdot {\bf S}$ spin-orbit interaction that split the orbitals into a lower $j = 1/2$ doublet and a $j = 3/2$ quadruplet that is higher in energy by $3\Lambda_{\rm so}$.  Crystal field effects arising from the graphene environment lead to $m$-dependence of $\epsilon_{|m|}$ and allow $\Lambda_{\rm so}'$ to differ from $\Lambda_{\rm so}$.  
Note also that we have assumed that the adatom's $s$-orbitals (and all other inner levels) are far in energy from the Dirac points and can be safely neglected.  This is indeed justified by our supercell density functional theory (DFT) calculations for both indium and thallium, where the 5$s$ and 6$s$ electrons contribute significantly to the density of states only at several eV below the Dirac points.  We have further ignored four-fermion interaction terms in $H_a$.  This, too, is justified for indium and thallium, since both elements donate their outer $p$ electron to graphene and form non-magnetic configurations.  In a perturbative tight-binding picture, electrons hopping onto the adatom from graphene therefore lead to a \emph{singly} occupied $p$ orbital, for which Coulomb repulsion is expected to be unimportant.  Finally, we note that $H_a$ is even under $R_z$ symmetry even though we did not enforce this.  In other words, the fact that the adatom is displaced away from the graphene sheet leads to no additional terms in the Hamiltonian $H_a$ compared to the (fictitious) case for an H-position adatom residing directly in the $(x,y)$ plane.  This fact is crucial for the irrelevance of induced Rashba terms for graphene, as we discuss further below.    

For the graphene-only part of the Hamiltonian, we posit that the usual nearest-neighbor hopping model $H_t$ is modified primarily through a change in chemical potential for the six sites adjacent to the adatom:
\begin{eqnarray}
  H_g &=& H_t -\delta \mu\sum_{j = 1}^6 c^\dagger_{{\bf r}_j}c_{{\bf r}_j}.
\end{eqnarray}  
Clearly $H_g$ is invariant under not only $R_{\pi/3}$, $R_x$, and $\mathcal{T}$, but also $R_z$ despite this symmetry being broken by the adatom.  (Terms which violate $R_z$ necessarily involve spin-orbit coupling, which we have justifiably neglected because it is exceedingly weak.)  

Since the adatom generically modifies the bond lengths between carbon atoms in its vicinity, the hopping amplitudes will be modulated as well near the adatom.  According to our DFT calculations, however, perturbations to graphene's lattice structure are quite small (the carbon bond lengths in the immediate vicinity of either indium or thallium change by less than 1\%).  We thus expect this effect to be minor and have neglected it for simplicity.  We have also neglected in $H_g$ any spin-orbit terms which directly couple the carbon atoms, which is orders of magnitude weaker than the spin-orbit interaction indirectly mediated by the adatom.  

\begin{figure}
\includegraphics[width = 8cm]{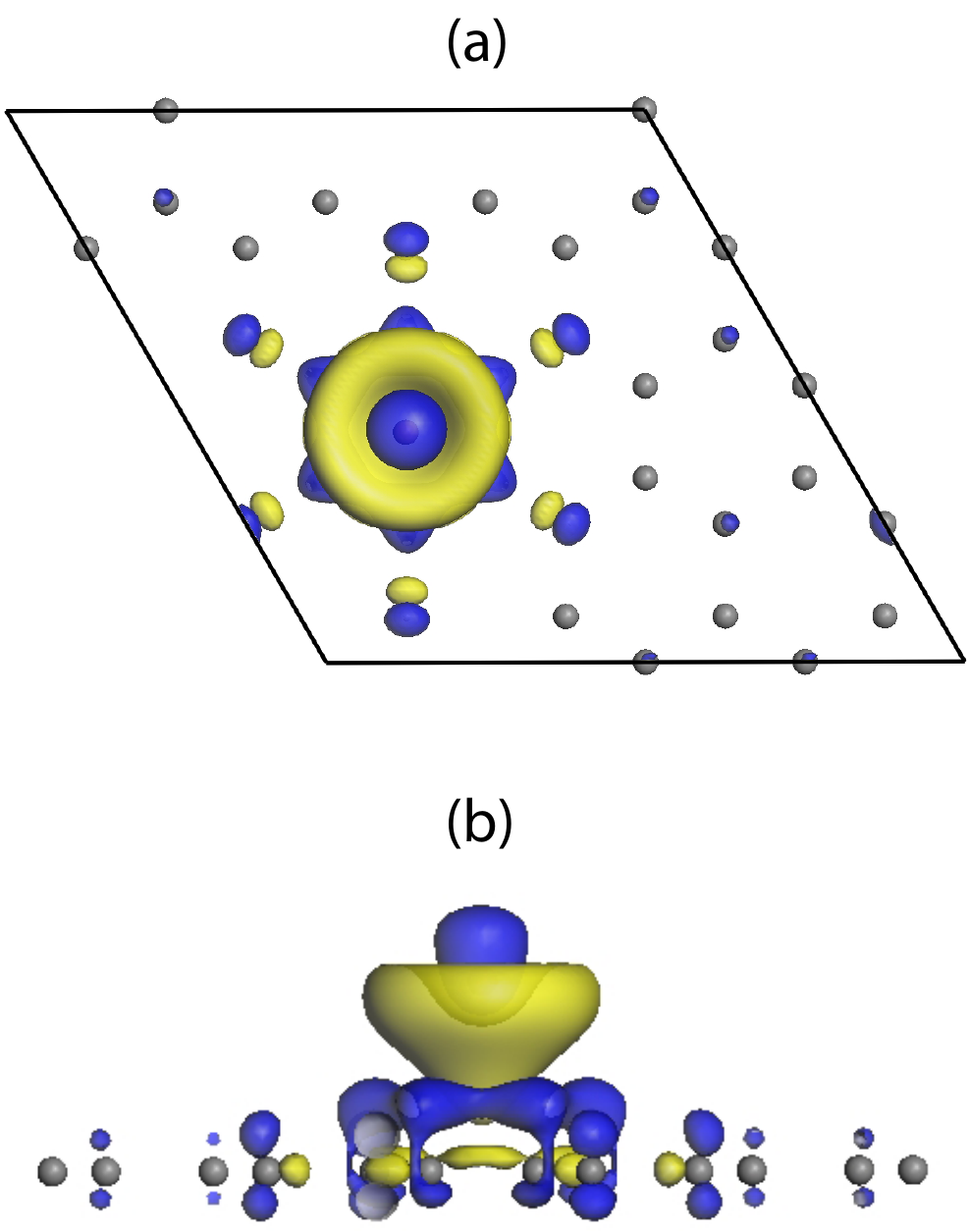}
\caption{{\bf Adatom-induced charge modulation.}  (a) Top and (b) side views of the charge density induced by a thallium adatom in a $4\times 4$ supercell.  Yellow/blue surfaces correspond to a positive/negative induced charge density (relative to the case where graphene and thallium decouple completely).  The charge modulations relax rather rapidly away from the adatom and occur mostly within the two innermost `rings' of carbon sites near the adatom.  }
\label{figS1}
\end{figure}

In contrast, the induced chemical potential $\delta \mu$ is not a weak effect---we find that our tight-binding simulations best reproduce the DFT band structure when $\delta \mu$ is on the order of 1eV.  Physically, $\delta \mu$ appears because the indium or thallium adatoms donate their outer $p$-orbital electron to the graphene sheet, leaving behind a net positive charge.  Electrons in the graphene sheet thus prefer to conglomerate in its vicinity in order to screen the positively charged adatom.  We note that of course there is no fundamental reason why the induced chemical potential should be confined to only the nearest six sites to the adatom.  Our DFT simulations demonstrate, however, that the induced charge modulation occurs very locally around the adatom, so that this is expected to be a good approximation.  As an illustration, consider the geometry with one thallium adatom in a $4\times 4$ supercell.  Figure \ref{figS1} displays the induced charge density $\Delta \rho \equiv \rho({\rm graphene + thallium})-\rho({\rm graphene})-\rho({\rm thallium})$ relative to the charge densities obtained when graphene and thallium decouple completely.  Here the yellow/blue regions correspond to isosurfaces with $\Delta\rho = \pm 0.01$eV/\AA$^3$.  Clearly the charge modulation occurs predominantly within the first two `rings' of carbon sites surrounding the adatom.    

Finally, let us discuss the Hamiltonian $H_c$ that allows electrons to tunnel between the adatom and graphene sheet.  For simplicity, we only allow for tunneling events which couple the adatom to its six nearest carbon atoms, and assume (because of carbon's weak intrinsic spin-orbit coupling) that such processes are spin-independent.  As in the main text it is useful to introduce linear combinations
\begin{equation}
    C_{m\alpha} = \frac{1}{\sqrt{6}}\sum_{j = 1}^6 e^{-i\frac{\pi}{3}m(j-1)} c_{{\bf r}_j\alpha}
\end{equation}
that, like $d_{m\alpha}$, carry angular momentum quantum number $m$ under $R_{\pi/3}$.  It follows from the symmetry transformation rules for $c_{\bf r}$ obtained above that $C_{m\alpha}$ transforms as follows:
\begin{eqnarray}
  R_{\pi/3}: ~~~~~~~~~~ C_m &\rightarrow& e^{i\frac{\pi}{6}s^z}e^{i\frac{\pi}{3}m}C_m
  \\
  R_x: ~~~~~~~~~~ C_m &\rightarrow& e^{i\frac{\pi}{2}s^x}C_{-m}
  \\
  {\mathcal T}: ~~~~~~~~~~ C_m &\rightarrow& e^{i\frac{\pi}{2}s^y} C_{-m}
  \\
  R_z: ~~~~~~~~~~ C_m &\rightarrow& -e^{i\frac{\pi}{2}s^z}C_m
\end{eqnarray}
Using the first three symmetries along with the corresponding transformation rules for $d_m$, it is straightforward to show that the spin-independent hoppings of interest must take the form
\begin{eqnarray}
  H_c = -\sum_{m = 0,\pm 1} (t_{|m|} C_{m}^\dagger d_{m} + h.c.),
\end{eqnarray}
with $t_0$ real and $t_1$ pure imaginary.  Under $R_z$ symmetry, the $t_0$ hopping is even while the $t_1$ hopping terms are odd.  This is easy to understand physically by considering the fictitious case where the adatom resides exactly in the graphene plane so that $R_z$ symmetry \emph{is} present.  In this case the $t_1$ terms reflect hopping between the adatom's $p_{x,y}$ orbitals and the carbon $p_z$ orbitals, which are orthogonal in this pathological limit.  That is, tunneling between graphene's $p_z$ orbitals and the adatom's $m = \pm 1$ orbitals arises only when $R_z$ is lifted.  In contrast, the adatom's $p_z$ orbitals overlap nontrivially with carbon's even when $R_z$ is preserved.  

Remarkably, the only terms in $H$ that do not respect $R_z$ are precisely these $t_1$ couplings.  To see what kind of spin-orbit processes the adatom can mediate, let us now consider the coupling between graphene and the adatom from a perturbative perspective.  An electron from the graphene sheet can tunnel onto the adatom via either $t_0$ or $t_1$.  Due to the adatom's strong spin-orbit coupling, the electron need not return to the graphene sheet via the same tunneling process by which it arrived.  For example, an electron can tunnel via the $t_0$ hopping, flip its spin, and then return to the graphene sheet via $t_1$.  Consequently, at second order in perturbation theory, the adatom will mediate effective couplings between the carbon atoms that are proportional to $t_0^2, t_1^2$, and $t_0 t_1$.  The electron's $S^z$ spin component is conserved in the former two cases, but not the third.  Furthermore, since the $t_0$ coupling in $H_c$ is even under $R_z$ while the $t_1$ coupling is odd, it is only the induced terms for graphene proportional to $t_0 t_1$ that violate $R_z$.  It follows that such spin-flip processes mediate \emph{Rashba} spin-orbit coupling, while the $t_1^2$ events mediate \emph{intrinsic} spin-orbit interactions.  (The $t_0^2$ processes do not generate spin-orbit coupling.)  It is worth noting as an aside that our ability to classify such processes in this simple way relies on the fact that all terms in $H_a$ are even under $R_z$.  If $H_a$ also involved terms that violated $R_z$, then \emph{any} second-order tunneling event would generically be capable of mediating Rashba spin-orbit coupling.  

As discussed in the main text, to generate a quantum spin Hall (QSH) state in graphene, the physics near the Dirac points must be dominated by intrinsic rather than Rashba spin-orbit interactions.  An important feature of our proposal is that Rashba coupling mediated by an adatom with a partially filled $p$ shell is always subdominant to the induced intrinsic spin-orbit terms.  This follows because electrons with momentum near the Dirac points hop very ineffectively onto the adatom's $m = 0$ orbital.  Previously we discussed how this effect arises because electrons destructively interfere when tunneling onto this orbital.  Here we provide a complementary perspective using the language of low-energy continuum Dirac fields.  Focusing on physics near the Dirac points, one can express the lattice fermion operators for the A and B sublattices of graphene as follows,
\begin{eqnarray}
  c_{{\bf r} \in A,\alpha} &\sim& e^{i {\bf Q}\cdot {\bf r}}\psi_{R,A\alpha} + e^{-i {\bf Q}\cdot{\bf r}} \psi_{L,A\alpha}
  \\
  c_{{\bf r} \in B,\alpha} &\sim& e^{i {\bf Q}\cdot {\bf r}}\psi_{R,B\alpha} + e^{-i {\bf Q}\cdot{\bf r}} \psi_{L,B\alpha}.
\end{eqnarray}
Here $\psi_{R/L\alpha}$ are slowly varying, two-component fields that describe low-energy excitations in the vicinity of the Dirac points at momenta $\pm {\bf Q} = \pm \frac{4\pi}{3\sqrt{3}a}{\bf \hat{x}}$ ($a$ is the separation between neighboring carbon atoms).  Let us use this decomposition to rewrite the operator $C_{m = 0}$ that appears in the $t_0$ tunneling.  Neglecting pieces involving derivatives of $\psi_{R/L\alpha}$, we obtain
\begin{eqnarray}
  C_{m = 0} &\sim& \frac{1}{\sqrt{6}}\bigg{[}\left(e^{i {\bf Q}\cdot {\bf r}_2} + e^{i {\bf Q}\cdot {\bf r}_4} + e^{i {\bf Q}\cdot {\bf r}_6}\right) \psi_{R,A} 
  \nonumber \\
  &+& \left(e^{i {\bf Q}\cdot {\bf r}_1} + e^{i {\bf Q}\cdot {\bf r}_3} + e^{i {\bf Q}\cdot {\bf r}_5}\right) \psi_{R,B}
  \nonumber \\
  &+& \left(e^{-i {\bf Q}\cdot {\bf r}_2} + e^{-i {\bf Q}\cdot {\bf r}_4} + e^{-i {\bf Q}\cdot {\bf r}_6}\right) \psi_{L,A} 
  \nonumber \\
  &+& \left(e^{-i {\bf Q}\cdot {\bf r}_1} + e^{-i {\bf Q}\cdot {\bf r}_3} + e^{-i {\bf Q}\cdot {\bf r}_5}\right) \psi_{L,B}\bigg{]}.
\end{eqnarray}
Remarkably, all terms in parenthesis above vanish identically.  The leading terms in the low-energy expansion of $C_{m = 0}$ thus involve derivatives and become, consequently, increasingly unimportant as one approaches the Dirac points.  

This implies that the Rashba coupling mediated by the adatom---which necessarily involves tunneling via $t_0$---is highly ineffective at influencing the physics at the Dirac points.  In contrast, $C_{m = \pm 1}$ admits a non-trivial expansion in terms of $\psi_{R/L\alpha}$ even when terms involving derivatives are dropped.  It follows that electrons with momenta near the Dirac points can effectively tunnel onto and off of the adatom via $t_1$, and the intrinsic spin-orbit coupling mediated by these hoppings thus always dominates over the induced Rashba terms.  As noted in the main text, this is \emph{not} necessarily the case for adatoms with partially filled $d$ or $f$ shells, because there are then additional orbitals onto which electrons from graphene can tunnel.  For instance, here electrons can tunnel onto the adatom's $m = \pm 2$ orbitals via a tunneling $t_2$; such processes, unlike $t_0$ tunneling, are \emph{not} suppressed at the Dirac points.  Effective couplings between carbon atoms proportional to $t_1 t_2$ are odd under $R_z$, involve spin flips, and correspond to Rashba spin-orbit interactions that may significantly affect the low-energy physics at the Dirac points.  The importance of such terms compared to the induced intrinsic spin-orbit interactions then depends on details of the Hamiltonian.  If, say, the additional orbitals hybridize very weakly with graphene or reside far in energy from the Dirac points due to crystal field effects, then $d$- or $f$-shell adatoms may still be suitable for stabilizing a QSH state in graphene.

\section{Density functional theory}

All DFT calculations were carried out with the Vienna ab-initio simulation package (VASP) \cite{Vasp1, Vasp2}, at the level of the
generalized-gradient approximation (GGA) \cite{PBE}. We used the projector augmented wave (PAW) method for the description of the
core-valence interaction \cite{Vasp3, Vasp4}. A vacuum space of 15{\AA} was employed in all calculations, along with a $15\times15$
$k$-grid mesh\cite{Monkhorst} for integrals in the two-dimensional Brillouin zone. The energy cutoff of the plane wave expansion was set to
500eV.

For pure graphene, the optimized lattice constant is 2.468{\AA}, consistent with the experimental value of 2.46{\AA}.  When addressing the
preferred location (H, B, or T) of different elements on graphene, a $4\times4$ supercell was employed with one adatom per cell. Recall
that we have two minimal criteria for stabilizing a QSH state in graphene: the adatoms should (i) prefer to occupy the H position and (ii) not form a net spin moment.  Among the elements we searched in this work, In, I, La, Hf, Au, Hg and Tl satisfy the first criterion, while the remaining elements
prefer either the T or B site. However, I, La, Hf and Au form magnetic moments and thus fail the second criterion, while the $s$ and $d$ bands of Hg reside far below graphene's Fermi level, implying a negligible interaction between the two. The only two remaining
elements---indium and thallium---satisfy both criteria and enter the H site non-magnetically. 

Including spin-orbit coupling, the
binding energies [$E_b = E({\rm adatom+graphene})-E({\rm graphene})-E({\rm atom})$] of indium and thallium on graphene at the H site are -0.525 and -0.133 eV,
respectively. The total energy per site when indium resides at the T and B sites is higher by 87 and 80meV, respectively, than when indium
resides at the H site. This is qualitatively consistent with the results of Ref.\ \onlinecite{CohenAdatoms} where spin-orbit coupling was
neglected. The conclusion is the same for thallium, though the preference for the H position is weaker; here the T and B sites are higher
in energy by 32 and 22meV. At low temperatures compared to these energy differences indium and thallium are expected to occupy only the H
position, with indium-carbon and thallium-carbon bond lengths of 2.83 and 2.90{\AA} as determined by GGA calculations.  We also calculated
atomic structures of indium and thallium on graphene within the local density approximation (LDA). The indium-carbon and thallium-carbon
bonds are shortened by only 0.09 and 0.06{\AA}, respectively, compared with the GGA results. It is thus reasonable to conclude that both GGA
and LDA can describe the structural and electronic properties well.

\section{Adiabatic continuity of the adatom and Kane-Mele models}  
  
We now turn to a periodic system with one indium or thallium adatom in a $4\times 4$ supercell.  This problem was explored extensively in the main text using both DFT and tight-binding simulations.  Our first-principles calculations demonstrated that, with either element, the Dirac points remain massless in the non-relativistic limit but acquire a substantial gap upon including spin-orbit coupling.  We further showed that the band structure and local density of states at the adatom site could be well-accounted for by an effective tight-binding model $H_{4\times 4}$.  This model included first-, second-, and third-neighbor hopping amongst carbon atoms in the graphene sheet, as well as the additional couplings discussed in the previous section around each adatom.  Note that inclusion of these further-neighbor tunnelings is necessary for obtaining reasonable quantitative agreement with first-principles band structure at energies away from the Dirac points, even for pure graphene\cite{GrapheneTightBinding}.  Since the adatoms' local density of states is concentrated at energies greater than 1eV above the Dirac points (see Fig.\ 2 of the main text), one needs these additional hoppings to accurately model their hybridization with the graphene bands.  (We stress that this is important for a quantitative description only.  None of the qualitative features discussed in this paper depend on these fine details.  For instance, with only nearest-neighbor hopping a gap still opens at the Dirac points, though the local density of states for the adatoms' $p_z$ states is difficult to accurately fit in this case.)  

\begin{figure}
\includegraphics[width = 6cm]{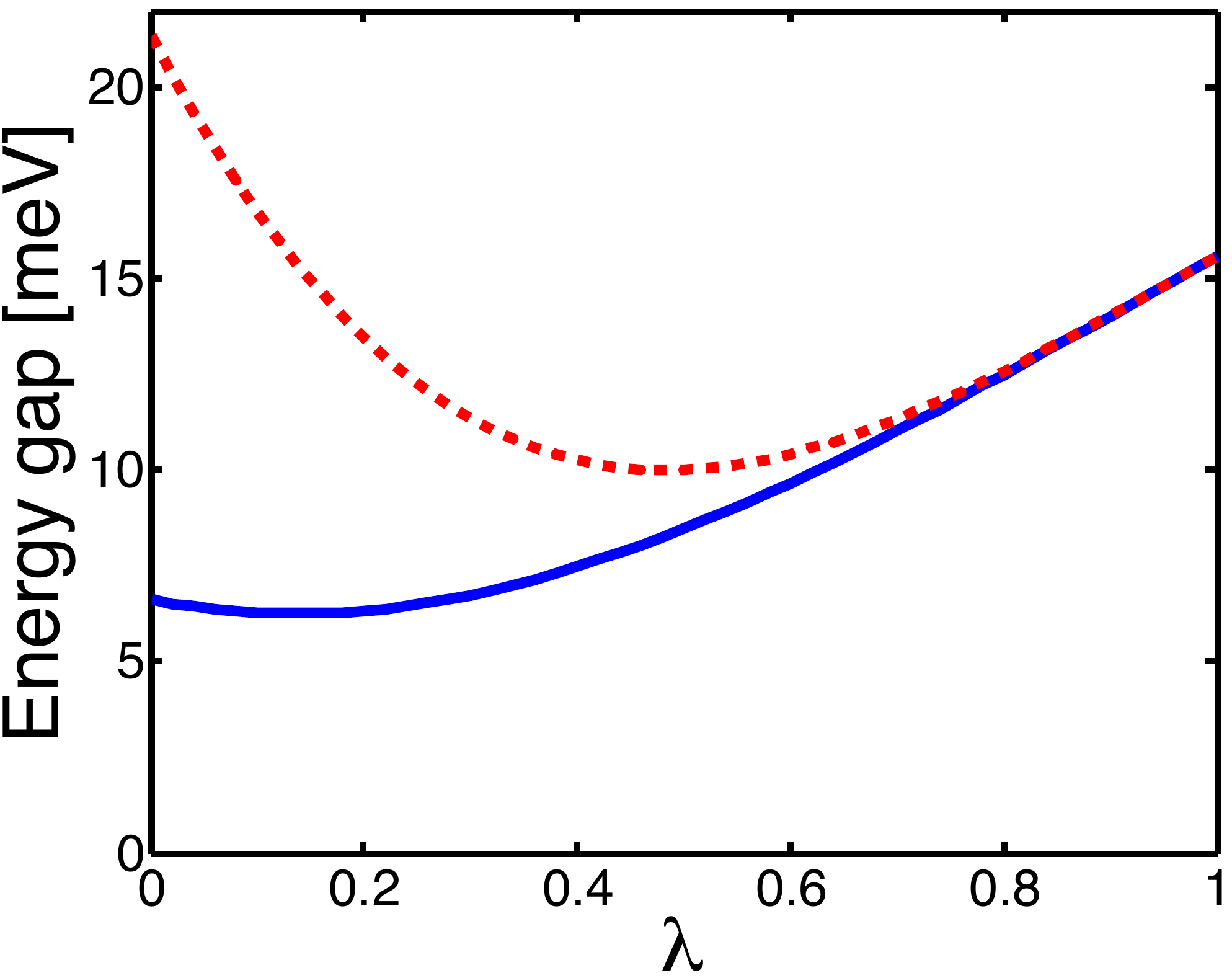}
\caption{{\bf Adiabatic continuity.}  Energy gap at the Dirac point as the Hamiltonian adiabatically deforms from the periodic adatom Hamiltonian $H_{4\times 4}$ at $\lambda = 0$ to the Kane-Mele model at $\lambda = 1$.  The solid blue and dashed red lines correspond to $H_{4\times 4}$ evaluated with parameters appropriate for indium and thallium, respectively.  In both cases the gap remains finite as $\lambda$ varies from 0 to 1, indicating that the Kane-Mele model and adatom Hamiltonian are adiabatically connected and thus support the same quantum spin Hall phase.}
\label{figS2}
\end{figure}  

One advantage of this effective tight-binding model is that it enables one to readily verify that the insulating regime generated by the spin-orbit-induced gap indeed corresponds to a QSH state.  This can be perhaps most easily proven by demonstrating that $H_{4\times 4}$ and the Kane-Mele model $H_{KM}$ [see Eqs.\ (1) and (2) of the main text] can be smoothly deformed into one another without closing the Dirac point gap.  To this end, consider the Hamiltonian $H(\lambda) = (1-\lambda)H_{4\times 4} + \lambda H_{KM}$ introduced in the main text that interpolates between these models as $\lambda$ varies from 0 to 1.  For $H_{KM}$ (which includes only nearest-neighbor hopping and uniform intrinsic spin-orbit coupling for graphene), we choose $t = 2.7$eV and $\lambda_{\rm so} = -0.0015$eV.  As Fig.\ \ref{figS2} illustrates, the Dirac point gap computed for $H(\lambda)$ indeed remains finite for all $\lambda$ between 0 to 1 with either indium (solid blue line) or thallium (dashed red line) parameters input into $H_{4\times 4}$.  The QSH state known to be supported by $H_{KM}$ and the insulating state stabilized by either type of adatom are, consequently, the same topological phase of matter.

\section{Graphene-only effective Hamiltonian}

In this section we outline the derivation of the graphene-only Hamiltonian, Eqs.\ (8-9) in the main text, from the tight binding model in Eqs.\  (4-7) describing the graphene sheet with H-position adatoms. We assume, for the purposes of this section only, that one adatom resides at each hexagonal plaquette but disallow any direct hopping of electrons between adatom $p$ orbitals. Under these assumptions we may exploit the translation symmetry of the problem and write the Hamiltonian (4) in momentum space as $H=\sum_\bk\psi_\bk^\dagger{\cal H}_\bk\psi_\bk$ with $\psi_\bk=(c^A_\bk,c^B_\bk;d_{1\bk},d_{0\bk},d_{-1\bk})^T$, where $c^{A/B}_\bk$ remove electrons from the A/B graphene sublattices while $d_{m\bk}$ remove electrons from the adatom orbitals. As before, spin is treated implicitly. The Hamiltonian matrix in this basis reads
\begin{equation}\label{hh1}
{\cal H}_\bk=\begin{pmatrix} h_g & T \\
T^\dagger & h_a
 \end{pmatrix},
\end{equation}
where
\begin{equation}\label{hhg}
h_g=\begin{pmatrix} 0 & \tau_\bk \\
\tau^*_\bk & 0
 \end{pmatrix}, \ \ \ \ \  \tau_\bk=-t\sum_{j=0}^2e^{i\bk\cdot\bd_j}
\end{equation}
is the standard nearest-neighbor tight-binding Hamiltonian for graphene with $\bd_j=(\sin{2\pi j\over 3},-\cos{2\pi j \over 3})$ unit vectors pointing from a site on sublattice A to its three neighbors on the B sublattice. Also,
\begin{equation}\label{hha}
h_a=\begin{pmatrix} 
\epsilon_1+\Lambda_{\rm so} s^z & \sqrt{2}\Lambda_{\rm so}' s^- & 0 \\
\sqrt{2}\Lambda_{\rm so}' s^+ & \epsilon_0 & \sqrt{2}\Lambda_{\rm so}' s^- \\
0 & \sqrt{2}\Lambda_{\rm so}' s^+ & \epsilon_1-\Lambda_{\rm so} s^z
 \end{pmatrix} 
\end{equation}
describes the adatom $p$ orbitals. Finally 
\begin{equation}\label{hhc}
T=\begin{pmatrix} 
V_1 & V_0 & V_{-1} \\
V^*_{-1} & V^*_0 & V^*_1 
 \end{pmatrix}, \ \ \ \ V_m=t_{|m|}\sum_{j=0}^2e^{i{2\pi j\over3}m}e^{-i\bk\cdot\bd_j}
\end{equation}
represents transitions between graphene and the adatom orbitals as described by $H_c$ in  Eq.\ (7).

We now wish to `integrate out' the adatom degrees of freedom and thus determine their effect on the electrons in graphene. The simplest way to accomplish this goal is to first perform a unitary transformation $\cH\to\tilde{\cH}=e^{-S}\cH e^S$ with $S$ chosen such that $\tilde{h}_g$ and  
$\tilde{h}_a$ in the transformed Hamiltonian are decoupled, i.e.\ $\tilde{T}=0$. In this basis the adatom degrees of freedom can be integrated out trivially. Following the steps outlined in Ref.\ \onlinecite{GrapheneSO3}
we find, to second order in $T$, 
\begin{equation}\label{hhg2}
\tilde{h}_g=h_g-{1\over 2}\left[T h_a^{-1}T^\dagger+h_gT h_a^{-2}T^\dagger +h.c.\right] +{\cal O}(T^4).
\end{equation}
This result is equivalent to treating Hamiltonian (\ref{hh1}) perturbatively to second order in $T$. 

We are interested in the effect of adatoms on the low-energy fermonic modes in graphene occuring in the vicinity of the Dirac points $\pm{\bf Q}$. In this limit, clearly, the second term in the brackets becomes negligible (since $h_g$ vanishes as $\bk\to\pm{\bf Q}$), so we therefore focus on the first term. Although it is possible to find an exact inverse of $h_a$ the result is cumbersome and tends to obscure the simple physics underlying the formation of the spin-orbit-induced gap in the system. To avoid this complication we make an additional assumption that spin-orbit splitting of the adatom orbitals is small compared to the crystal field effects, $|\Lambda_{\rm so}|,|\Lambda_{\rm so}'|\ll |\epsilon_m|$, and write $h_a=h_0+h_1$ with $h_0={\rm diag}(\epsilon_1,\epsilon_0,\epsilon_1)$. To first order in $\Lambda_{\rm so}$ and $\Lambda_{\rm so}'$ we then have  $h_a^{-1}\simeq h_0^{-1}-h_0^{-1}h_1 h_0^{-1}$ and 
\begin{equation}\label{hhg3}
\tilde{h}_g\simeq h_g-T h_0^{-1}T^\dagger+T h_0^{-1}h_1h_0^{-1}T^\dagger.
\end{equation}
The adatom-induced correction to $h_g$ is seen to have a simple structure. The first term is spin-independent and mediates on-site potential terms and hoppings between sites in the hexagon surrounding an adatom. The second, spin-dependent term in turn produces the intrinsic and Rashba spin-orbit couplings.    
\begin{figure*}
\includegraphics[width = 18cm]{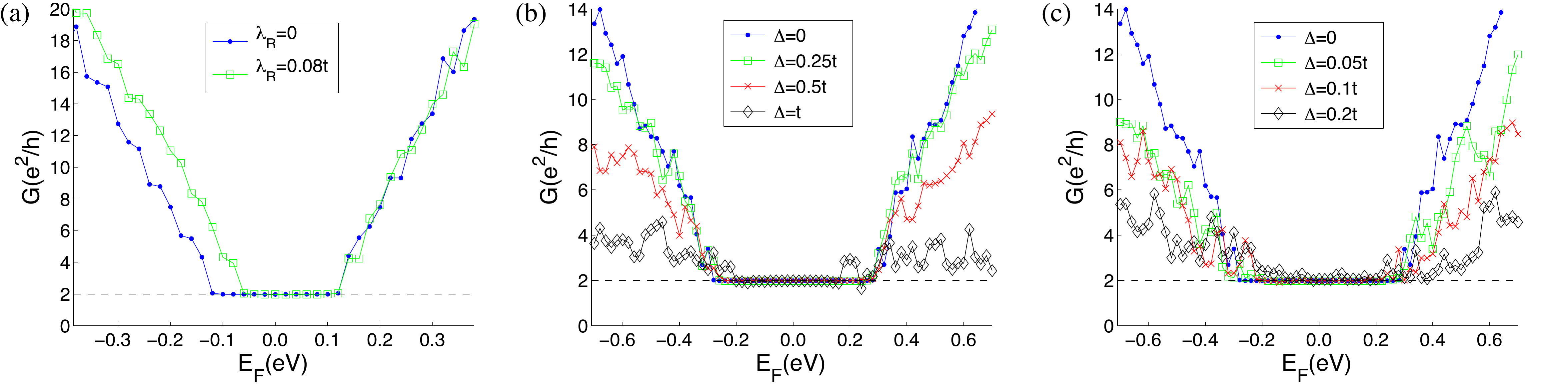}
\caption{{\bf Hexagonal Rashba effect and substrate-induced disorder.}  Conductance $G$ as a
function of the Fermi energy $E_F$ for a system including (a) hexagonal Rashba coupling, (b) uncorrelated disorder randomly distributed over the range $\left[-\Delta, \Delta\right]$ and (c) correlated disorder as defined in Eq.\ (\ref{gauss}) with $\rho=0.07$ and $d=5$.  
In all cases the system is of size $W=80$ and $L=40$ with adatom coverage $n_i=0.2$ and nearest-neighbor hopping strength $t=2.7$eV. For panel (a) the intrinsic spin-orbit strength is $\lambda_{\rm so}=0.04t$; for panels (b) and (c), $\lambda_{\rm so}=0.1t$.
}
\label{figS3}
\end{figure*}

Various terms resulting from Eq.\ (\ref{hhg3}) can be explicitly worked out in a somewhat tedious but straightforward calculation. The spin-orbit terms take the form  
\begin{equation}\label{dhhg}
\delta h_g^{\rm SOC}=\begin{pmatrix} R_1+D & R_2 \\
R_2^\dagger & R_1-D
 \end{pmatrix},
\end{equation}
with
\begin{equation}\label{Dress1}
D=\Lambda_{\rm so}{|t_1|^2\over \sqrt{3}\epsilon_1^2}s^z\sum_{j =0}^2\sin{\bk\cdot(\bd_j-\bd_{j+1})}
\end{equation}
and  
\begin{eqnarray}
R_1 &=&
\Lambda_{\rm so}'{\sqrt{2}t_0|t_1|\over 3\epsilon_0\epsilon_1}
\sum_{j,l}[{\bf s}\times(\bd_l-\bd_j)]_z\sin{\bk\cdot(\bd_j-\bd_l)}, \nonumber\\
R_2 &=& -i
\Lambda_{\rm so}'{\sqrt{2}t_0|t_1|\over 3\epsilon_0\epsilon_1}
\sum_{j,l}[{\bf s}\times(\bd_l+\bd_j)]_z e^{-i\bk\cdot(\bd_j+\bd_l)}. \nonumber
\end{eqnarray}
If we now recall that $\bd_0+\bd_1+\bd_2=0$ and identify 
\begin{equation}\label{coupl}
\lambda_{\rm so}=\Lambda_{\rm so}{|t_1|^2\over 2\sqrt{3}\epsilon_1^2}, \ \ \ \ 
\lambda_R=\Lambda_{\rm so}'{\sqrt{2}t_0|t_1|\over 3\epsilon_0\epsilon_1}
\end{equation}
we see that $D$ and $R_{1,2}$ correspond respectively to the intrinsic and hexagon Rashba spin-orbit couplings in Eq.\ (9). One can readily verify that $R_{1,2}$ vanish at $\bk=\pm {\bf Q}$ while $D$ approaches a nonzero value of $3\sqrt{3}\lambda_{\rm so}$, as expected on the basis of arguments presented above.

Using the tight-binding parameters obtained in Sec.\ III of the main text we may estimate $\lambda_{\rm so} \simeq 23$meV and $\lambda_R=58$meV for thallium (for indium both are about factor of 3 smaller). At 6\% coverage, assuming uniform averaging of $\lambda_{\rm so}$ over all plaquettes, this would imply a spin-orbit-induced gap of  $0.06\times 6\sqrt{3}\lambda_{\rm so}\simeq 14$meV, somewhat smaller that the $21$meV gap predicted by DFT at the same coverage. We attribute this discrepancy to the neglect of higher-order terms in Eq.\ (\ref{hhg3}); since $t_{0,1}$ are of similar magnitude as $\epsilon_{0,1}$ this expansion cannot be expected to yield quantitatively accurate values of the coupling constants. Nevertheless the procedure outlined above is useful in that it illustrates how various symmetry-allowed spin-orbit terms emerge in the effective graphene-only model. It also confirms that although the Rashba coupling exceeds the intrinsic coupling by more than a factor of two, its special form, dictated by symmetry, forces it to vanish at the Dirac point and renders $\lambda_R$ irrelevant for the low-energy physics.  

Using the same methods one can also estimate various spin-independent terms induced by the adatoms. These include the on-site potential and additional hopping terms mentioned above. Since the latter are already present in pristine graphene and since adatom corrections are generally small compared to these bare values, we expect their effect on the low-energy physics to be minimal and do not discuss them further.

\section{Hexagonal Rashba and Disorder effects}

As mentioned in Sec.\ IV of the main text, the stability of the QSH phase was also tested
against the addition of hexagonal Rashba coupling and residual disorder in the graphene sheet.  We now discuss these results in greater detail.  Throughout this section we will consider random adatom arrangements at coverage $n_i = 0.2$, modeled using the graphene-only Hamiltonian from Eq.\ (8) of the main text with $\delta \mu = 0$ for simplicity.  We first consider the influence of the hexagon Rashba term defined in Eq.\ (9). While it is true that its effects identically vanish at the Dirac points, it nevertheless causes some splitting of the bands in their vicinity.  As a result these Rashba terms can reduce somewhat the size of the mobility gap observed with either periodic or random adatoms, though to a far lesser extent than the traditional Rashba coupling. The splitting will depend on the ratio of $\lambda_{\rm so}$ to $\lambda_R$, and, more importantly, the absolute strength of $\lambda_{R}$ compared to $t$.

In Figure \ref{figS3}(a) we show the result for a set of parameters chosen to illustrate this effect while avoiding the complications due to the finite size of the system. Here $\lambda_{\rm so}=0.04t$ and $\lambda_R=0.08t$, their ratio chosen to be similar to that estimated in Eq.\ (\ref{coupl}) above. The mobility gap has clearly been somewhat reduced, but a robust plateau remains. We remark that the standard Rashba term (with only nearest neighbor coupling) at the same coupling strength would completely destroy the plateau. For smaller values of $\lambda_R$ and $\lambda_{\rm so}$ (but keeping their ratio constant) the effect of hexagonal Rashba coupling becomes even less pronounced although the finite size effects prevent us from obtaining reliable results at low adatom coverage in this regime.  It is also worth emphasizing that one should \emph{not} view $\lambda_R$ as reducing the gap compared to those found in our DFT and tight-binding results displayed in Fig.\ 2 of the main text.  Those simulations incorporated both the intrinsic \emph{and} hexagon Rashba spin-orbit couplings (among other couplings neglected for simplicity here), and the results already reflect the influence of the latter.

To study residual disorder unrelated to the adatoms (\emph{e.g.}, arising from impurities in the substrate), two models were used, both of which can be captured by an on-site potential term of the form specified in Eq.\ (8) of the main text.  In the first model, we incorporate an uncorrelated random chemical potential in the 
range $\left[-\Delta, \Delta\right]$ on every lattice site.  In the second, we employ a longer-range, correlated disorder potential of the form\cite{Yuan}
\begin{equation} \label{gauss}
\delta\mu_{{\bf r}_{i}}=\sum_{j=1}^{N_{\rm imp}}V_{j}\exp{\left(-\frac{\left|{\bf r}_{i}-{\bf r}_{j}\right|^{2}}{2d^{2}}\right)},
\end{equation}
where $N_{\rm imp}$ is the number of impurity sites, $V_{j}$ is also uniformly random over the range $\left[-\Delta, \Delta\right]$,
and $d$ is the radius of the impurity potential. As in Ref.\ \onlinecite{Yuan}, we characterize $N_{\rm imp}$ by the ratio $\rho=N_{\rm imp}/N$, with  $N$ being the total number of carbon atoms in the sample.

While we have not averaged over multiple disorder realizations as in Fig.\ 3(b) of the main text, with only a single disorder realization one can still see
the general trend towards instability of the QSH phase as $\Delta$ increases. Our simulations show that this state, despite already being stabilized by a disordered arrangement of adatoms, in fact exhibits remarkable resilience against additional uncorrelated substrate-induced disorder in the graphene sheet.  As Fig.\ \ref{figS3}(b) illustrates, in this case a very high value of $\Delta\approx t$ can be applied before the first signs of disorder affecting the topological phase appear, and even here a considerable conductance plateau survives.  The topological phase is, however, more sensitive to long-range correlated disorder, which is likely more relevant for experiment when charged impurities from the substrate present the main disordering mechanism.  Figure \ref{figS3}(c) illustrates that here
the deterioration of the plateau is more rapid. At $\Delta=0.2t\approx 0.54$eV, with $\rho=0.07$ and $d=5$, there is still a remnant of the gapless edge states, but for stronger disorder the topological phase is destroyed.  Notice that this value of $\Delta$ is comparable to the width of the plateau in the `clean' case where the only disorder source is the randomly distributed adatoms.


\end{document}